\documentclass[conference]{IEEEtran}
\IEEEoverridecommandlockouts
\usepackage{cite}
\usepackage{algorithmic}
\usepackage{xcolor}
\usepackage{enumerate}
\usepackage{amsfonts,amsmath,amssymb,amsfonts}
\usepackage{amsthm}
\usepackage{algorithmic}
\usepackage{optidef}
\usepackage{algorithm}
\usepackage{mathrsfs}
\usepackage{graphicx}
\usepackage{textcomp}
\usepackage{subfigure}
\usepackage[squaren]{SIunits}
\usepackage[section]{placeins}

\def\BibTeX{{\rm B\kern-.05em{\sc i\kern-.025em b}\kern-.08em
    T\kern-.1667em\lower.7ex\hbox{E}\kern-.125emX}}
\begin{document}
\newtheorem{theorem}{\textbf{Theorem}}
\newtheorem{lemma}{\textbf{Lemma}}
\bibliographystyle{IEEEtran}

\title{Parameter Estimation based Automatic Modulation Recognition for Radio Frequency Signal}
\author{\IEEEauthorblockN{Shuo Wang\IEEEauthorrefmark{1}, Kuojun Yang\IEEEauthorrefmark{1}, Zelin Ji\IEEEauthorrefmark{2}, Qinchuan Zhang\IEEEauthorrefmark{1}, Huiqing Pan\IEEEauthorrefmark{1}}
\IEEEauthorblockA{\IEEEauthorrefmark{1}School of Automation Engineering, University of
 Electronic Science and Technology of China, China\\
\IEEEauthorrefmark{2} Shenzhen Institute for Advanced Study, University
 of Electronic Science and Technology of China, China.\\
Email: 202211060911@std.uestc.edu.cn, zelinji77@gmail.com, \{yangkuojun,
zhangqc, panhuiqing\}@uestc.edu.cn}
}
\maketitle

\begin{abstract}
Automatic modulation recognition (AMR) critically contributes to spectrum sensing, dynamic spectrum access, and intelligent communications in cognitive radio systems. The introduction of deep learning has greatly improved the accuracy of AMR. However, current automatic identification methods require the input of key parameters such as the carrier frequency, which is necessary to convert the radio frequency (RF) to a base-band signal before it can be used for identification. In addition, the high complexity of deep learning models leads to high computational effort and long recognition times of existing methods, which are difficult to implement in demodulation system deployments. To address the above issues, in this paper, we first use power spectrum analysis to estimate the carrier frequency and signal bandwidth, which realizes the effective conversion from RF signals to base-band signals. This paper chooses the long short-term memory (LSTM) network as the model for automatic identification, which has low implementation complexity while maintaining high accuracy. Finally, by training the LSTM with actual sampling data combined with parameter estimation (PE), the method proposed in this paper can guarantee more than 90\% format recognition accuracy. 
\end{abstract}


\section{Introduction}

In the field of modern wireless communications, cognitive radio systems significantly improve spectrum utilisation and communication efficiency through spectrum dynamic access~\cite{8233654} and automatic modulation recognition (AMR) techniques~\cite{8645696, 10531791, 10570601}. AMR enables the system to intelligently sense and identify signal modulation types in the radio spectrum, identify under-utilised spectrum resources, and achieve dynamic spectrum access (DSA)~\cite{9463441}. In addition, it supports cognitive radio to dynamically adjust transmit parameters, such as modulation type and power, to adapt to the current spectrum environment and avoid interference to the primary user~\cite{10304321, 10683785}. AMR plays a critical role, thus enhancing communication quality, system performance, and spectrum usage flexibility. 

Machine learning (ML) have emerged as a solution for AMR with a view to exploiting their superior performance in data processing and intelligent recognition. However, this approach relies heavily on expert experience to extract features and select appropriate classifiers. Typical classifiers include decision tree (DT)~\cite{9776529}, k-nearest neighbor (KNN)~\cite{8885738} and support vector machine (SVM)~\cite{6573231,8333735}. However, ML-based AMR methods can lead to erroneous discrimination once the assumed signal model or noise model does not match the actual signal or noise.

Researchers have applied deep learning to cognitive radio systems~\cite{mfrl2023ji}, especially for AMR tasks~\cite{9293316, 8977561, 9188007}. Meng~\textit{et al.}~\cite{8454504} propose an end-to-end convolutional neural network (CNN)-based AMR (CNN-AMR) that outperforms feature-based methods and is closer to the optimal ML-AMR. In addition, the CNN-AMR is robust in estimating errors of carrier phase offset and signal-to-noise ratio (SNR). Parallel CNN transformer network (PCTNet) is also used for AMR, which is a model based on CNN's advantage of extracting local information, adding the advantage of capturing long-distance dependencies~\cite{10478085}. Deep residual neural networks (ResNet)~\cite{9194036} are designed to extract high-level feature representations for excellent classification accuracy. To extract the modulated signal special features in multiple dimensions, Chang~\textit{et al.}~\cite{9462447} proposed a deep neural network (MLDNN) based on multi-task learning, which effectively fuses both I/Q and A/P dimensional features. However, the above deep learning based techniques mainly focus on the classification of the base-band signals, which requires precise signal parameters, e.g., the carrier frequency, for the demodulation. However, in certain blind demodulation scenarios, the key parameters are unknown and can only be estimated, which may lead to mismatch of the signal type and erroneous discrimination.

To address the above issues, we proposed PE based AMR (PE-AMR), to achieve the AMR with lower complexity, as well as implement the proposed PE-AMR algorithm in the practical signal demodulation scenarios. However, to implement the proposed PE-AMR algorithm, we face several challenges:
\begin{itemize}

\item {Existing AMR methods can only identify baseband signals; however, blind demodulation systems, such as IoT anti-attack systems, have no knowledge of all modulation parameters, rendering existing AMR methods unusable.}

\item {Deploying deep learning in demodulation systems based on acquisition platforms. The recognition speed is too slow to meet the demand of real-time demodulation, and the resource consumption is too large for demodulation systems to bear.}

\end{itemize}

Therefore, to improve the utility and reliability of automatic format recognition, there is an urgent need to develop novel demodulation methods that reduce both the dependence on user input and the computational complexity. The contributions of this paper address the above challenges and are concluded as follows.

\begin{enumerate}
\item {\textbf{Demodulation parameter estimation (PE)}: Power spectrum analysis is used to estimate the carrier frequency and signal bandwidth to realize the effective conversion of radio frequency (RF) signal to baseband signal and facilitate the work of AMR module.}

\item{\textbf{Long short-term memory (LSTM) network integration}: LSTM network is chosen as the model for AMR, which maintains high accuracy with low implementation complexity.}

\end{enumerate}


\section{System Model}
\label{chp6:system}

\begin{figure}[t]
\centering
\newpage
\includegraphics[width=0.9\columnwidth]{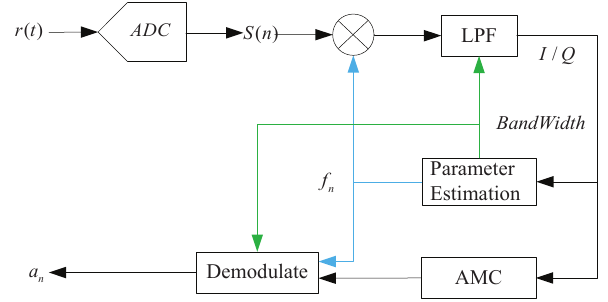}
\caption{The proposed demodulation model.}
\label{system_model}
\end{figure}

Fig.~\ref{system_model} shows the demodulation system model proposed in this paper.The input to the system is RF signals, and the system can classify and demodulate RF signals without requiring user input of carrier and bandwidth and modulation format.\par 

The system is divided into three parts. The first part is the RF base-band signal conversion (RBC) module which converts the RF input to base-band signal. The second part consists of an AMR module and a PE module, which outputs the estimated parameters to the mixing and demodulation modules. The PE module needs to output the estimated carrier frequency and signal bandwidth to the RBC module. And output the signal oversampling rate (i.e., the number of sampling points in one symbol period) to the demodulation module. The third part is the demodulation module, which accomplishes the demodulation of the signal according to the modulation parameters and modulation type.

The mathematical model of the system is presented next. The formula for RF signals is as follows
\begin{equation}
   r(t) = \sum_{n=-\infty}^\infty{a_{n} g_{T}(t-nT_{b})} \cdot 
   e^{j2\pi f_{c}t} + n(t),
\label{fum_rt}
\end{equation}

\noindent where $a_{n}$ is base-band transmission information. $T_{b}$ is the symbolic period and $g_{T}(t)$ is the unit impulse response of the pulse molding filter. We assume that the pulse molding filter is the root raised cosine roll down filter, and its roll down factor is $\rho$, $0<\rho<1$. $n(t)$ is additive white Gaussian noise, and the power spectral density of one side is $\frac{N_{0}}{2}$.\par

The symbol rate $R_s$ of the signal $r(t)$ is related to the signal bandwidth $B$
\begin{equation}
   {B} = (1+\rho) \cdot {R_s},
\label{fum_baudrate}
\end{equation}

\noindent where ${B}$ is the signal bandwidth and ${R_s}$ is the symbol rate.

Observing Fig.\ref{system_model}, the RF signal is sampled at equal time intervals by an analog-to-digital converter (ADC) and will be changed into a sampled signal (i.e., discrete signal). The sampling value of $r(t)$ at moment point $kT_s$ is

\begin{equation}
   r(kT_s) = \left[\sum_{n=-\infty}^\infty{a_{n} g_{T}(t-nT_{b})} \cdot 
   e^{j2\pi f_{c}t} + n(t)\right]_{t=kT_s}.
\label{fum_rkts}
\end{equation}

The $k$-th sampling value of $r(t)$ is

\begin{equation}
   r(k) = \sum_{n=-\infty}^\infty{a_n g_{\rm{T}}(kT_{\rm{s}}-nT_{\rm{b}}-\epsilon T_{\rm{b}})} \cdot e^{j2\pi f_{\rm{c}}kT{\rm{s}}} + n(k),
\label{fum_rn}
\end{equation}

\noindent where $\epsilon T_{{b}}$ is a timing error, and $\left| \epsilon \right|<1$, is caused by different clock sources at the sending module and the receiving module. Meanwhile, $T_{{s}}$ is the receiving sampling rate, $n(k):=n(t)|_{{t=nT_{{s}}}}$. Assuming a sampling rate of $f_{{s}}$, we can be obtain $sps$ as follows
\begin{equation}
   sps = \frac{f_{{s}}}{R_s},
\label{sps}
\end{equation}

\noindent where $sps$ is the oversampling rate, the PE module needs to feed it to demodulation module, which can be obtained from the $B$.\par

The mixer and the low-pass filter require two parameters, the carrier frequency and the $B$ of the signal. Equation~(\ref{fum_xk}) can be obtained from~(\ref{fum_rn}) after adding frequency errors.
\begin{equation}
    \begin{aligned}
           x(k) = \sum_{n=-\infty}^\infty{[a_n g_{\rm{T}}(kT_{\rm{s}}-nT_{\rm{b}}-\epsilon T_{\rm{b}})} \cdot 
           \\
           e^{j(2\pi f_{\rm{e}}kT{\rm{s}}+\theta_{\rm{e}})}] + n(k),
    \end{aligned}
\label{fum_xk}
\end{equation}

\noindent where $f_{\rm{e}}$, $\theta_{\rm{e}}$ represents carrier frequency offset and carrier phase offset, which is caused by the difference between the carrier of the converter module and the transmitter module, and $f_{\rm{e}}<\frac{1}{T_{\rm{b}}}$, $\left|\theta_{\rm{e}}\right|<\pi$. The different carrier frequencies of the mixing and transmitting modules, the ones in the cooperative system are due to the Doppler effect. The system of this paper it is due to the error of PE module. There is a contradiction between the AMR module and the PE module. When the accuracy of PE is high, the frequency deviation of the input signal of the AMR module is small, and when the accuracy of PE is low, the frequency deviation of the input signal of the AMR module is large. 

\section{Parameter Estimation}
\label{algorithm}
The parameters to be estimated are the carrier frequency and the $B$ of the RF signal. They are used for mixer and low-pass filter. The estimation errors of these can be solved by the training process of the AMR module. The power spectrum combined with the smoothed average periodogram method is used for PE.

At the end of the nineteenth century, the German scholar Schuster first proposed a method of spectral estimation, named periodogram method. The basic principle is: For the length of $L$ signal to do analysis, first according to the length of N will be divided into equal lengths of multiple segments, and then through the signal segmentation spectrum averaging to obtain an estimate of the power spectrum of the signal power spectrum after the transformation to get the power spectrum of the channel for the (\ref{P_n}).

\begin{equation}
   P[n] = \sum_{k=1}^{L/N} \frac{\left| FFT \left\{ x[(k-1)N+n] \right\} \right|^2}{N},
\label{P_n}
\end{equation}

\noindent where $FFT\left\{ \cdot\right\}$ is the fourier transform corresponding to the signal x[n]. Assuming that $x[n],0\leq n \leq N-1$ for a random signal, the length of N, then its Fourier transform for the (\ref{FFT}).

\begin{equation}
   X[k] = \sum_{n=0}^{N-1} {x[n]e^{-j\frac{2\pi n}{N}} }.
\label{FFT}
\end{equation}

Direct FFT, due to the truncation of the signal, will cause spectral leakage, the general solution is to add a window. For the window function, rectangular window, Hamming window, Hanning window, Blackman window and so on are more commonly used. Each of the above four window functions has its own characteristics, and in practical engineering applications, it is necessary to choose the appropriate window function according to the design objectives. Here in this paper, the Blackman window is chosen, and the mathematical expression of its window function is (\ref{FFT_windows}).

\begin{equation}
   \omega (n)=
    \begin{cases}
    \begin{array}{l}
        -0.5 \cos{\frac{2 \pi n}{N-1}} +0.42 + 
        \\0.08 \cos{\frac{4 \pi n}{N-1}}  
    \end{array}
    & 0\leq n \leq N-1,\\ 
    \begin{array}{l}
        0 
    \end{array}
    & \text{otherwise}.

    \end{cases}
\label{FFT_windows}
\end{equation}

\begin{figure}[t]
    \subfigure[Power spectrum.]{
        \begin{minipage}[t]{0.5\linewidth}
            \centering
            \includegraphics[width=\columnwidth]{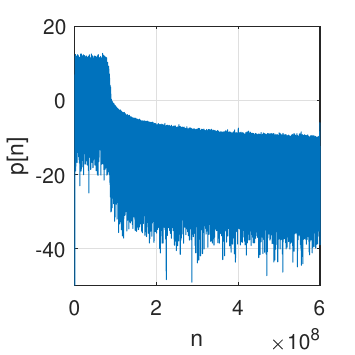}
            \label{power_spectrum}
        \end{minipage}%
    }%
    \subfigure[Histogram corresponding.]{
    \begin{minipage}[t]{0.5\linewidth}
        \includegraphics[width=\columnwidth]{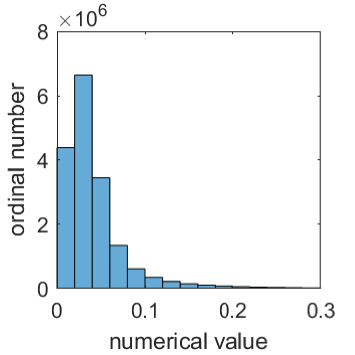}
        \label{bar_chart}
    \end{minipage}
}%

\caption{Power spectrum and the histogram corresponding of QAM16 signal.}
\end{figure}

In order to realize the estimation of the signal bandwidth and carrier frequency, the signal power spectrum needs to be analyzed. The power spectrum of a QAM16 signal is shown in Fig.~\ref{power_spectrum}, where the power spectral density of Gaussian white noise is a constant value approximating the noise floor of $P_{v}$. The value of $P_{v}$ can be obtained by histogram analysis of the power spectrum.

As shown in Fig.~\ref{bar_chart}, the mid-value of the interval with the highest probability of occurrence in the histogram is the estimation of $P_{v}$. The value of $P_{v}$ can be obtained by histogram analysis of the power spectrum.

Find the interval (assuming a total of $M$ points) where the value of the ordinal n in the power spectrum is greater than $\frac{P_{v}+P_{max}}{2}$, where $P_{max}$ is the maximum value in the power spectrum.

\begin{equation}
   \left\{n1,n2,\ldots,n_{M} \right\} = \mathop{find}\limits_{n \in [1,N]} \left\{ p[N]\geq \frac{P_{v}+P_{max}}{2} \right\}.
\label{P[n]}
\end{equation}

This gives an estimate of the signal bandwidth as

\begin{equation}
   B = \frac{f_{s} \cdot M}{N}.
\label{bandwidth}
\end{equation}

An estimate of the carrier frequency can be obtained by also taking the planting $N_{c} = \frac{n_{1} + n_{W}}{2}$ in this interval

\begin{equation}
   f_{c} = \frac{N_{c}-1} {N_{p}} \cdot f_{s}.
\label{carrier}
\end{equation}

\section{LSTM Model for Automatic Modulation Recognition}

\subsection{LSTM Model}
\label{troditional LSTM model}

\begin{figure}[t]
\centering
\newpage
\includegraphics[width=\columnwidth]{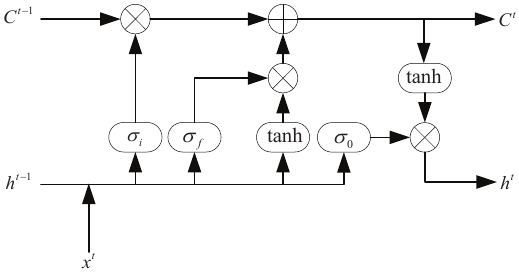}
\caption{Typical LSTM node.}
\label{LSTM_Node}
\end{figure}

A typical LSTM node is shown in Fig.\ref{LSTM_Node}. It has a short-term memory state $h^t$ and a long-term memory state $c^t$, where $c^t$ varies slowly from layer to layer, and $h^t$ varies greatly from node to node, directly determining the predicted output at that moment. Assuming that the time sequence of each input is $x^t$, with $a^t=[ x^t;h^{t-1}]$ as the sequence of operations this time, the splicing training to get four states. Respectively, with the $tanh$ activation function and the $sigmoid$ activation function $\sigma$ to do the nonlinear processing.\par

Input data:
\begin{equation}
   z = tanh(W \times a^t).
\label{Input_data}
\end{equation}

Forgetting parameter:
\begin{equation}
   z^f = \sigma (W^f \times a^t ).
\label{Forgetting_parameter}
\end{equation}

Memory paramete:
\begin{equation}
   z^i = \sigma (W^i \times a^t ).
\label{Memory_paramete}
\end{equation}

Output parameter:
\begin{equation}
    z^o=\sigma(W^o \times a^t ).
\label{Output_paramete}
\end{equation}

Where $W^f, W^i, W^o$, and $W$ are the weight matrices for forgetting, memorizing, output, and input, respectively. Meanwhile, $\sigma$ is defined as $\sigma(x)=1/(1+e^{-x})$, and $tanh(x)=(e^x-e^{-x})/(e^x+e^{-x})$. And $t$ is current moment, $t-1$ is previous moment. There are two main phases experienced within an LSTM neuron:\par
\begin{enumerate}[1)]
\item Forgetting and remembering. The retention level of the previous state $c^{t-1}$ and the admission level of the new input state $z$ are determined by the forgetting parameter $z^f$ and the remembering parameter $z^i$. Find their Hadamard products separately
\begin{equation}
    c^t = z^f \odot c^{t-1}+ z^i \odot z.
\label{Hadamard_products}
\end{equation}

\item Synthesize the output vector. Determine the short-term memory state $h^t$ from the long-time memory state $c^t$ of this node and the output parameter $z^o$ as
\begin{equation}
    h^t = z^o \odot tanh(c^t ).
\label{short_term_memory_state}
\end{equation}
\item Output. The node output is determined by the sigmoid function as
\begin{equation}
    y^t = \sigma {(W^{'} h^t)}.
\label{Output}
\end{equation}

\end{enumerate}

\subsection{LSTM Implementation for AMR}
\label{LSTM implementationl}

\begin{figure}[h]
\centering
\newpage
\includegraphics[width=0.5\linewidth]{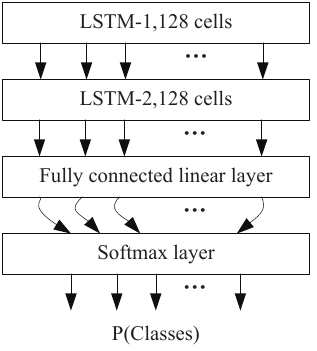}
\caption{LSTM model designed for AMR.}
\label{AMR_LSTM_Node}
\end{figure}

The RNN-based model has a better fitting ability for time-dependent sequences, and the LSTM further adds a forgetting mechanism. We designs a network architecture with two LSTM layers as shown in Fig.~\ref{AMR_LSTM_Node}, by extracting the feature quantity of the signal after inputting the dataset at the point of this simulation of the damage effect on the signal by the actual channel, and then subsequently the extracted features are trained to finally obtain the desired classifier Model. In the recognition model, the receiver receives the recognition signal from the network, and after feature extraction, the signal is recognized according to the LSTM model.

In the proposed PE-AMC, consists of two layers of LSTM cell, a fully connected (FC) layer and a softmax layer. The two layers of LSTM node are identical and the input size is 128$\times$2, the hidden layer has 128 neurons. We use the Adam optimiser for training PE-AMC and set the batch size to 400 and the learning rate to $1e^{-3}$.

\begin{figure}[t]
\centering
\newpage
\includegraphics[width=0.9\linewidth]{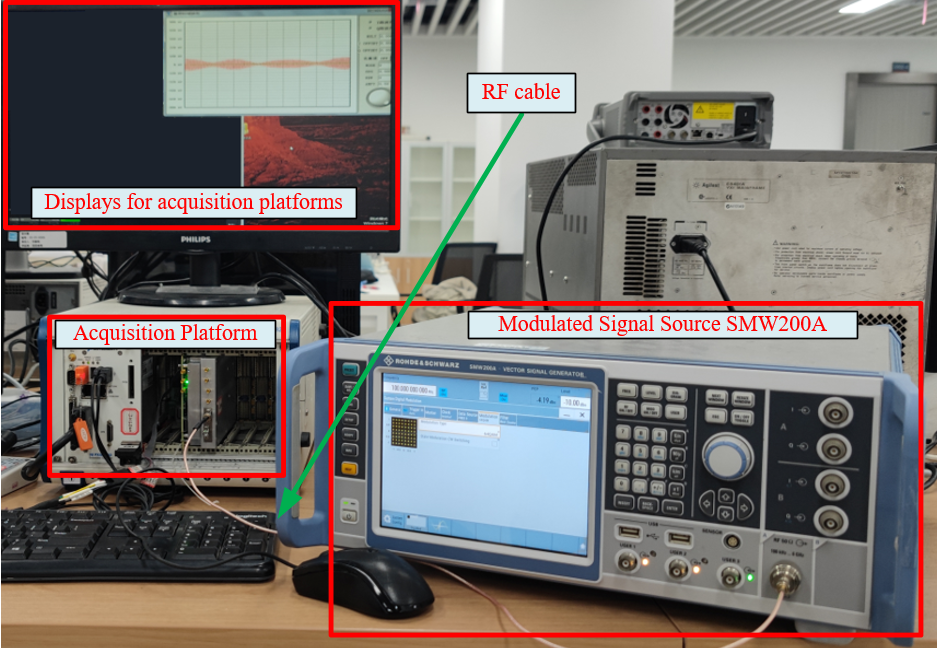}
\caption{Data acquisition platforms and digital modulated signal sources.}
\label{acq_system}
\end{figure}

\section{Experiments}

\label{Experiments}
The goal of PE-AMR is to find a network where the signaling errors due to PE can be resolved in the training phase, i.e., the model is more accurate and requires less complexity. Therefore, we do two experiments. Experiment I verifies the performance of the LSTM on the dataset of RML2016.a. Experiment II verifies the performance of LSTM after PE in the actual sampled dataset will bring signal error.

\subsection{Modulation Datasets}
\label{Modulation Datasets}

\subsubsection{RML2016.a}

To demonstrate the advantages of LSTM over other methods in terms of performance and computational complexity, this paper metaphysical public database, RML2016.a from deepsig. Which contains 20 different signal to noise ratios ranging from -20dB to 18dB at 2dB intervals, and for each signal to noise ratio there are also 11 different modulation formats, including 8 digital and 3 analog, so that the total number of different signal types is 220. Each modulation format contains 1000 groups of sample points, each sample point is in the format of (2,128), i.e., 128 data for the I and Q signals. This dataset is for channel simulation using dynamic channel model hierarchical blocks which are defined by frequency offsets, sampling rate offsets, AWGN, multipath and fading. Dataset RadioML.2016.10a's includes 11 modulations

\begin{itemize}
    \item Analog modulation: AM-DSB, AM-SSB and WBFM.
    \item Digital modulation: 8PSK, BPSK, CPFSK, GFSK, 4PAM, 16QAM, 64QAM and QPSK.
\end{itemize}

\begin{table}[t]
    \centering
    \caption{Dataset for the sampled signal} \label{real_data}
    \begin{tabular}{|c|c|}
        \hline
        \textbf{Parameter} & \textbf{Value} \\
        \hline
        {Carrier frequency $f_{c}$} & 1GHz\\
        \hline
        {Symbol rate $R_s$} & [200M, 210M, $\cdots$, 300M] \\
        \hline
        {Roll-off factor $\rho$} & [0.1, 0.2, $\cdots$, 0.9]\\
        \hline
        {Signal-to-noise ratio} & $\approx 25$ dB\\
        \hline
    \end{tabular}
 \end{table}

\subsubsection {Actual sampling dataset}
In order to verify the effectiveness of the PE-AMR, we utilizes a data acquisition platform to actually collect RF data and generate a dataset. The data sampling platform and digital modulated signal sources is shown in the Fig.\ref{acq_system}. Sample platform parameters: bandwidth is 8 GHz, sampling rate is 20 GHz. The digital modulation source is the SMW200A from ROHDE~\&~SCHWARZ. The signal source can generate a variety of standard modulated signals. The signal format includes BPSK, QPSK, 8PSK, 16QAM, 64QAM, 256QAM, and the rest parameters of the sampled signal are shown in {TABLE~\ref{real_data}}. After the mixer and low-pass filter, do 10 times decimation, the signal becomes a base-band signal with a sampling rate of 2GSPS.

\subsection{Performance Comparison for RML2016.a}
\label{Performance comparison based on RML2016.a}
\begin{figure}[t]
\centering
\newpage
\includegraphics[width=0.9\linewidth]{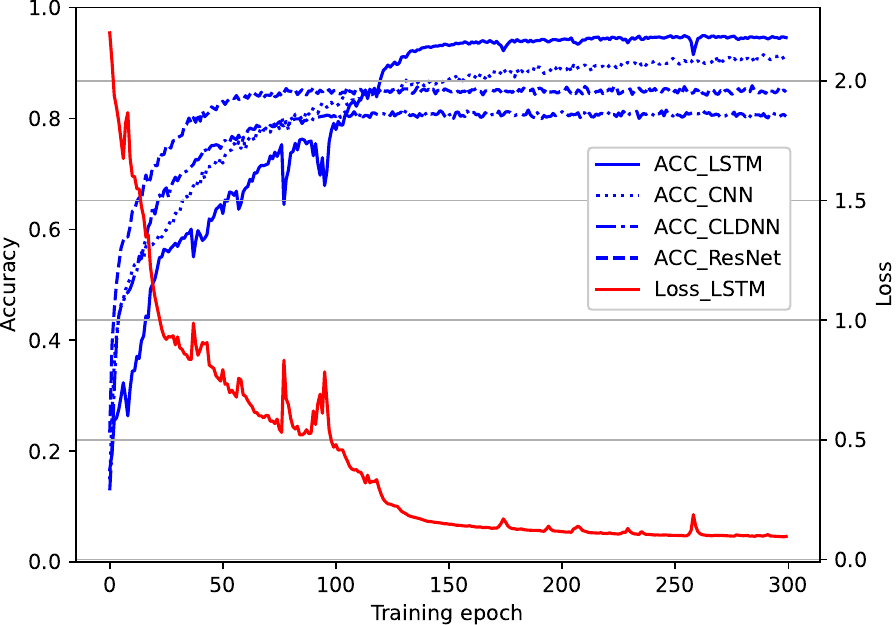}
\caption{LSTM-based modulation recognition accuracy.}
\label{Training_acc_loss_four_method}
\end{figure}

\begin{figure}[t]
\centering
\newpage
\includegraphics[width=0.9\linewidth]{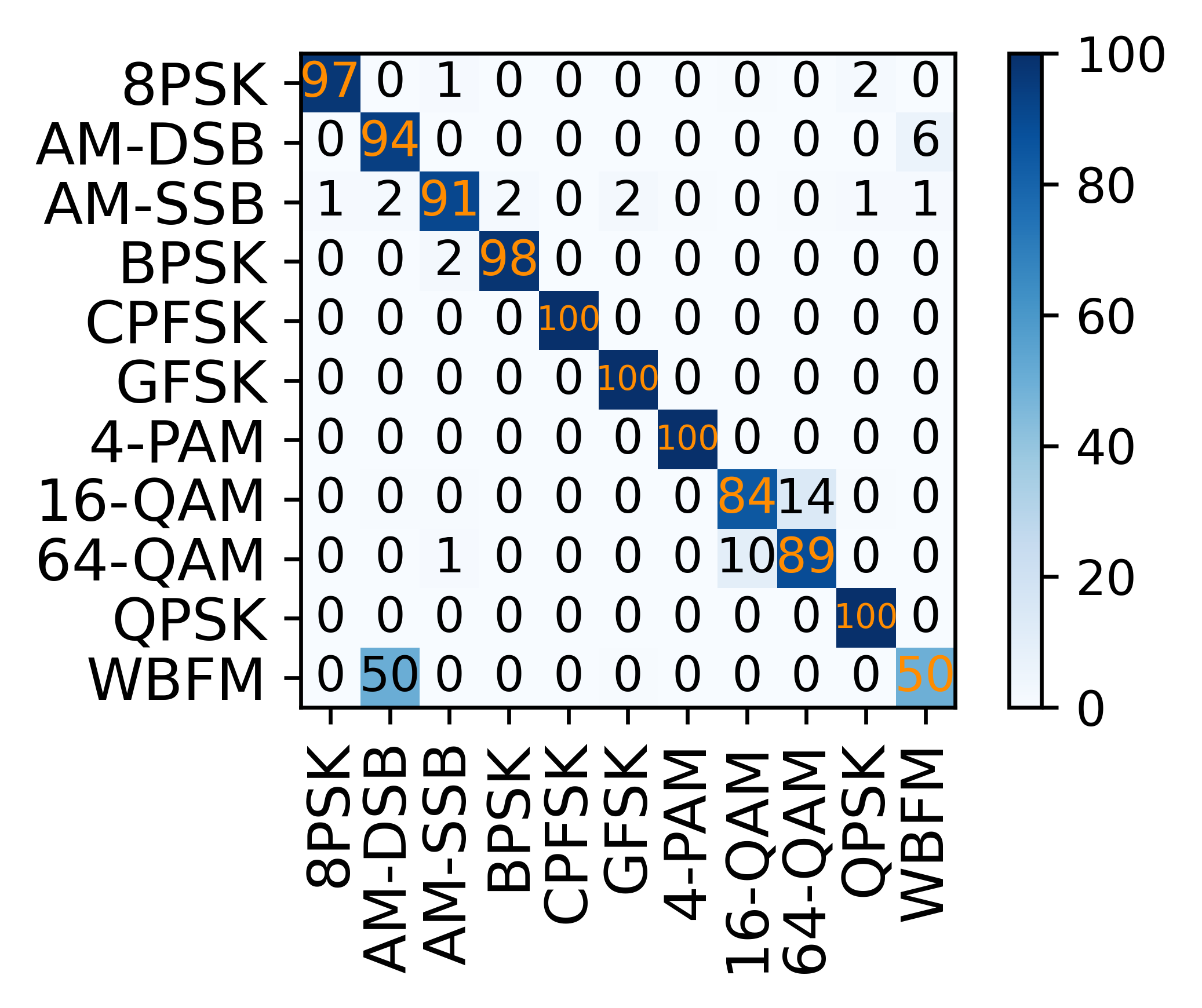}
\caption{LSTM-based confusion matrix.}
\label{confusion_matrix}
\end{figure}

We compare the LSTM scheme with other network structures include CNN1, ResNet, and CLDNN as benchmarks. By dividing the 220,000 data in 6:2:2 and randomly selecting 132,000 data as the training set and further disrupting it to increase the pervasiveness, by taking out the data with a SNR of 18dB for validation, the training accuracy is obtained as shown in Fig.~\ref{Training_acc_loss_four_method}. The accuracy of LSTM has surpassed that of other models after 150 generations of training. It can also be seen that the training accuracy and training loss curves move in opposite directions. To observe the model recognition accuracy in terms of different modulation formats, Fig.~\ref{confusion_matrix} shows the confusion matrix of the LSTM on 4dB data. The recognition accuracy is above 90\% for most formats.

\begin{figure}[t]
\centering
\newpage
\includegraphics[width=0.9\linewidth]{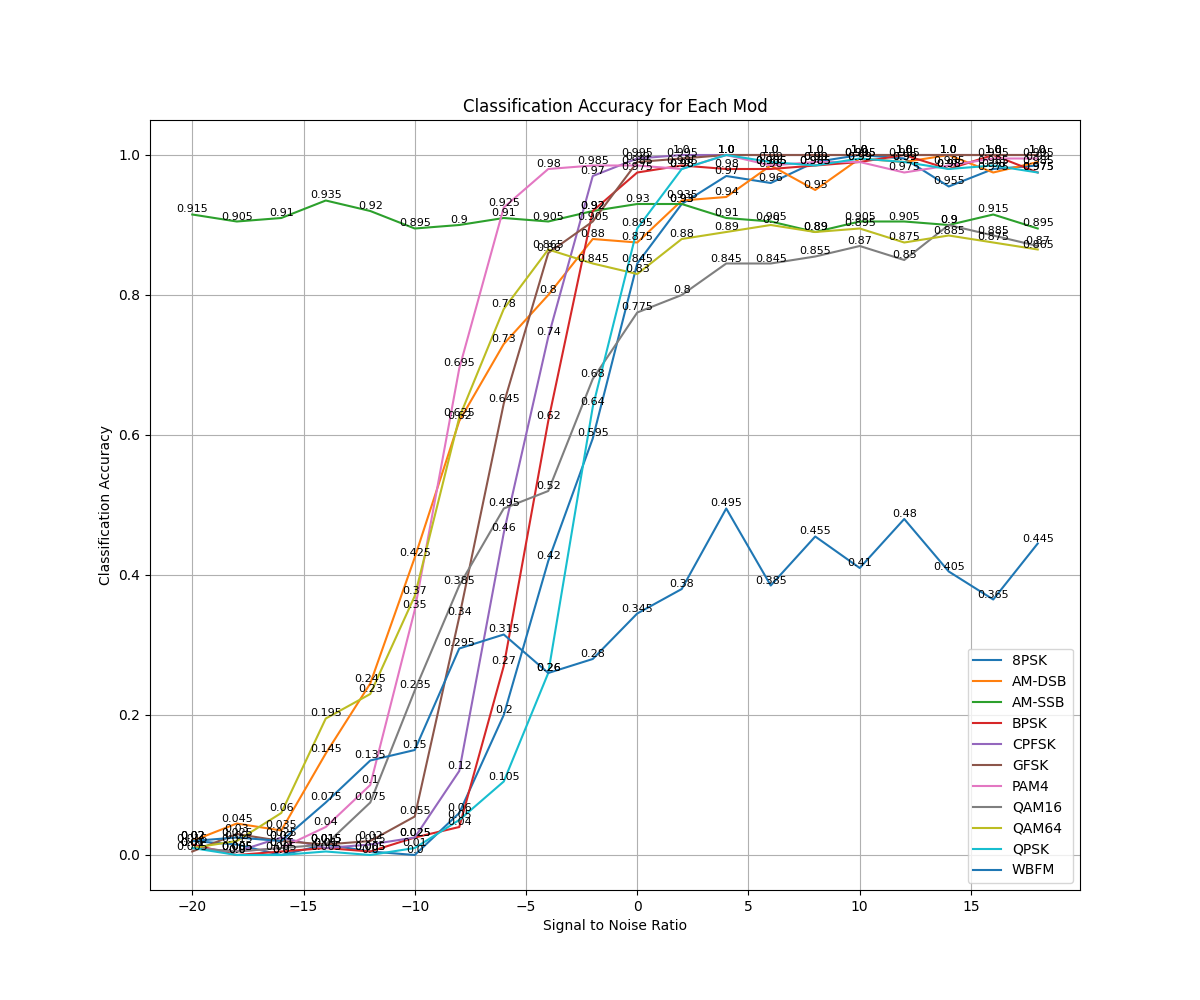}
\caption{LSTM-based modulation recognition accuracy.}
\label{LSTM_acc_snr}
\end{figure}

Further, we use the model's ability to adapt to the signal-to-noise ratio. As shown in Fig.~\ref{LSTM_acc_snr}, the data of 11 modulation formats at 20 signal-to-noise ratios are introduced into the trained two-layer LSTM neural network model, which can achieve high accuracy prediction of the modulated signal type. After the signal-to-noise ratio is greater than 4dB, the recognition accuracy of some modulation types can reach more than 90\%. Under general test conditions, the signal-to-noise ratio can reach about 30dB, which is much higher than 0dB, so the prediction using this model can achieve a high recognition accuracy, which can further promote the subsequent timing synchronization, carrier synchronization and equalization functions effectively. CNN1, ResNet, and CLDNN based on RML2016.a Recognition accuracy is not as good as LSTM.\par

We also investigate the execution time latency for the proposed LSTM scheme and other benchmarks, which are shown in Table~\ref{elapsed time for model}. The experimental conditions for the models here are all using the RXT4090, based on the tensorflow platform. The elapsed time here calculates the time for model startup, etc., but since the hardware and software platforms are the same, the rest of the elapsed time for these models is the same. From table~\ref{elapsed time for model}, we can get that LSTM has the shortest elapsed time, so it is considered that LSTM has the lowest computational complexity. Combined with the previous experiments, it can be concluded that LSTM has the highest accuracy and the lowest computational complexity.

\begin{table}[t]
    \centering
    \caption{Table of elapsed time for model AMR} \label{elapsed time for model}
    \begin{tabular}{c|c|r|c|c}
        \hline
        {Model} & {CNN1} & {ResNet} & {CLDNN} &  {LSTM} \\
        \hline
        {Elapsed time} & 1.4854 & 1.5505 & 1.6540 & 1.0848\\
        \hline
    \end{tabular}
 \end{table}

\subsection{Performance Analysis for Sampling data}
\label{Performance analysis based on Sampling data}

Finally, we verify the performance of the proposed PE-AMR scheme by the real sampling data. We build a database after obtaining the sampling data of RF signals using the acquisition platform. The signals in the database are parameterized to estimate their carrier frequency and signal bandwidth using the periodogram method combined with the power spectrum. According to the comparison between the estimation results and the real values, the accuracy of this method is within $10^{-3}$. And according to the PE structure to do mixing and low-pass filtering to get the baseband signal, to generate the signal identification database of the actual collected signals.\par

\begin{figure}[t]
\centering
\newpage
\includegraphics[width=0.8\linewidth]{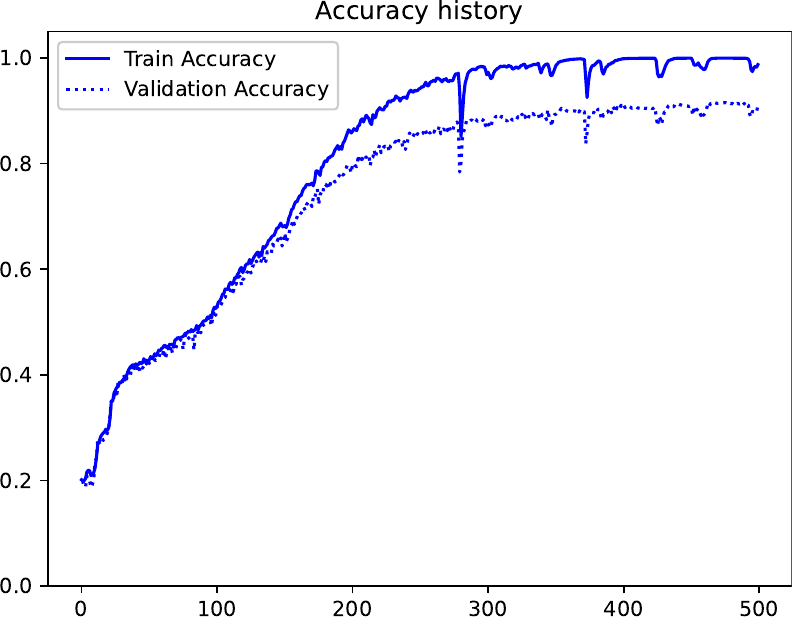}
\caption{Modulation recognition accuracy of real collected data based on LSTM.}
\label{accurancy}
\end{figure}

The 300,000 data in this database are divided in the ratio of 6:2:2 and 180,000 data are randomly selected as the training set. And further disrupt the training set to improve the generalization ability, and get the recognition accuracy graph during the training process, as shown in Fig.\ref{accurancy}. After 250 generations of training, the recognition accuracy has been greater than 90\%. From the Fig.\ref{accurancy}, it can be seen that the LSTM remains accurate in the actual sampled data with errors from PE.

\section{Conclusion}
\label{chp6:conclusion}

In this paper, PE and automatic format recognition have been combined to achieve high accuracy and low complexity automatic format recognition using LSTM model. The RF signal bandwidth and carrier frequency have been realized using the power spectrum combined with the periodogram method. We have applied the RML2016.a database to compare the four models CNN1, ResNet, LSTM and CLDNN, where LSTM has shown the lowest computational complexity and the highest recognition accuracy. Moreover, we have established the actual acquisition platform, combined with PE and baseband signal conversion module resume sampling signal database. The experimental results have proved that even if the PE brings extra errors, LSTM can still guarantee more than 90\% recognition accuracy.

\bibliography{Reference}
\clearpage
\end{document}